\documentclass{DISproc}

\usepackage{float}
\usepackage{epsfig}
\usepackage{amsmath}
\usepackage{amssymb}
\usepackage{eucal}
\usepackage{cite}
\usepackage[rflt]{floatflt}

\newcommand{\lsim}{\raisebox{-0.07cm   }
{$\, \stackrel{<}{{\scriptstyle\sim}}\, $}}

\begin{document}
%------------------------------------
\title{{\small 
       \textnormal{DESY-12-125~~       DO-TH 12/21~~
                   SFB/CPP-12-49~~LPN 12-079
}}\\
Higher Twist contributions to the Structure Functions \boldmath{$F_2(x,Q^2)$} and 
$g_2(x,Q^2)$}

%for single authors the superscripts are optional
\author{{\slshape Johannes Bl\"umlein$^1$, Helmut B\"ottcher$^1$}\\[1ex]
$^1$Deutsches Elektronen-Synchrotron, DESY, Platanenallee~6, D-15738 Zeuthen, Germany\\
}

% please enter the contribution ID for the DOI
\contribID{xxxxx}

\doi  % if there is an online version we will register DOIs

\maketitle

\begin{abstract}
\noindent
We report on recent results on higher twist contributions to the unpolarized structure functions
$F_2^{p,d}(x,Q^2)$ at N$^3$LO in the large $x$ region and constraints on the twist--3 contribution
to polarized structure function $g_2(x,Q^2)$.
\end{abstract}

%--------------------------------------------------------------------------------------------------
\section{Introduction}

Higher twist terms contribute to the nucleon structure functions at lower scales $Q^2$. The range
in which these terms may be safely neglected against the leading twist contributions, partly depends 
on the size of the experimental errors in the respective measurement. Highly precise data at low values 
of $Q^2$ allow to access these contributions, the detailed physical understanding of which is presently 
still in an early stage. It has been outlined in Refs.~\cite{BBG,Blumlein:2008kz} how the higher twist 
contributions can be extracted in a phenomenological way in case of the structure functions $F_2(x,Q^2)$ 
and $F_L(x,Q^2)$ in the valence quark region. In this note we report on recent results of an improved 
analysis. Another interesting question concerns the structure function $g_2(x,Q^2)$ 
in the polarized case, which has been measured to a higher precision during the last years \cite{DATA}. 
Here we try to extract first information on the twist-3 contributions to $g_2(x,Q^2)$.

%--------------------------------------------------------------------------------------------------
\section{Higher Twist Contributions to \boldmath $F_2^{p,d}(x,Q^2)$}
%--------------------------------------------------------------------------------------------------

We have carried out a QCD analysis in the valence region including more recent data from JLAB following 
earlier work \cite{BBG}. In the present analysis tails from sea-quarks and the gluon in the valence 
region were dealt with based on the ABKM distributions \cite{Alekhin:2009ni}. Both the valence quark 
distributions
$xu_v(x,Q^2_0)$ and $xd_v(x,Q^2_0)$ at $Q_0^2 = 4~\GeV^2$ are effected only very little. The values of 
$\alpha_s(M_Z^2)$ change marginally w.r.t. the earlier analysis  \cite{BBG}. We obtain~: $\alpha_s(M_Z^2) = 
0.1148  \pm 0.0019~\text{NLO},
= 0.1134  \pm 0.0020~\text{NNLO};~0.1141 \pm 0.0021~\text{N$^3$LO$^*$}$. Here, the N$^3$LO$^*$-analysis
accounts for the three-loop Wilson coefficients and a Pad\'e-model for the non-singlet four-loop anomalous 
dimension, 
to which 
we attached a $\pm 100 \%$ uncertainty, cf.~\cite{BBG} for details. Furthermore, we found that the response
of the individual deep-inelastic data sets in the valence region respond stable values, which are in 
accordance 
with the central value obtained moving from NLO to N$^3$LO$^*$. The present result agrees very well with
determinations of $\alpha_s(M_Z^2)$ in Refs.~\cite{Alekhin:2009ni,GJR,Alekhin:2012ig}, 
see also~\cite{Alekhin:2011ey}. A survey on the current status of $\alpha_s(M_Z^2)$ based on precision
measurements in different reactions has been given in \cite{Bethke:2011tr}. In the present analysis we
obtain a lower value of $\alpha_s$ than the world average, cf.~\cite{Bethke:2011tr}, and values being 
obtained in \cite{MSTW,NNPDF} at NNLO. Reasons for the difference to the values given in \cite{MSTW,NNPDF}
have been discussed in Refs.~\cite{Alekhin:2012ig,Alekhin:2011ey} in detail. In particular, the partial 
response of $\alpha_s$ in case of the BCDMS and SLAC data in \cite{MSTW,NNPDF} turns out to be partly different 
comparing to the results in \cite{Alekhin:2009ni,GJR,Alekhin:2012ig}. There are also differences between  
the analyses \cite{MSTW} and \cite{NNPDF} w.r.t. several data sets contributing.

The higher twist contributions can be determined by extrapolating the fit-results at leading twist
for $W^2 > 12.5~\GeV^2$ to the region $4 < W^2 < 12.5~\GeV^2, Q^2 \geq 4~\GeV^2$, 
cf.~\cite{Blumlein:2008kz,BB12a}.\footnote{In Ref. \cite{Alekhin:2012ig} also higher twist contributions for $x$ 
below the 
valence region have been determined.} The results for the coefficients $C_{\rm HT}^{p,d}(x)$ 
%--------------------------------------------------------------------------------------------------
\begin{eqnarray}
F_2(x,Q^2) = F_2(x,Q^2)\left[\frac{O^{\rm TM}[F_2(x,Q^2)]}{F_2(x,Q^2)} + \frac{C_{\rm HT}(x)}{Q^2[\GeV^2]}
\right]
\end{eqnarray}
%--------------------------------------------------------------------------------------------------
are shown in Figure~1, where we averaged over the respective range in $Q^2$. We applied the target mass 
corrections \cite{Georgi:1976ve} to the leading twist contributions.\footnote{An unfolding of the target mass 
corrections of the DIS world data for $F_2$ and $F_L$ including the JLAB data, has been performed in 
\cite{Christy:2012tk} recently.}  The result for the higher twist 
coefficients for proton and deuteron targets depends on the order to which the leading twist distribution is 
described. The higher twist terms become  smaller moving from NLO to N$^3$LO$^*$. Within the present theoretical 
and 
experimental accuracy the curves stabilize for $x < 0.65$, while at larger values there are still differences.
%----------------------------------------------------------------------------------------------
\restylefloat{figure}
\begin{figure}[H]
\begin{center}
\mbox{\epsfig{file=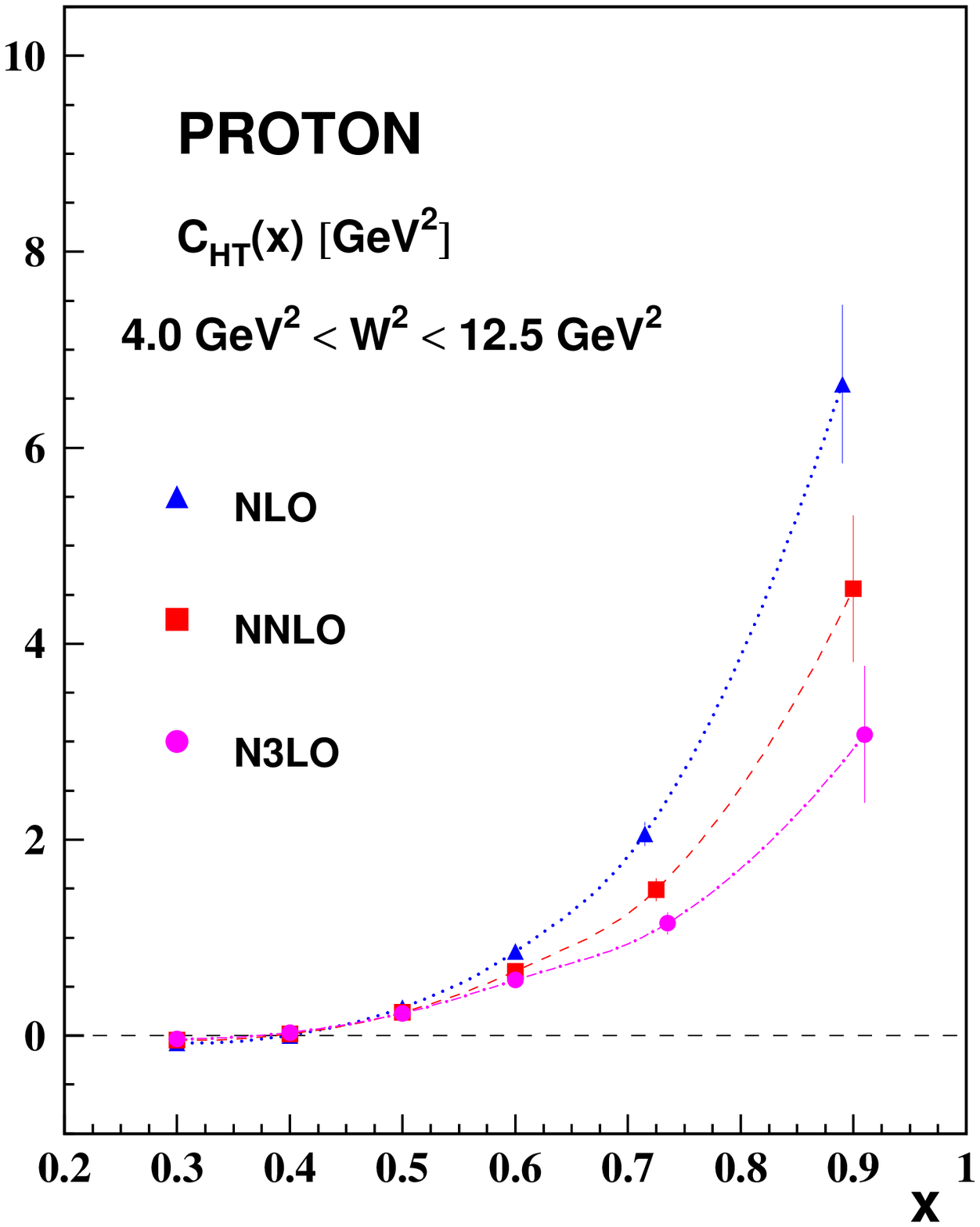,height=7cm,width=7cm}}
\mbox{\epsfig{file=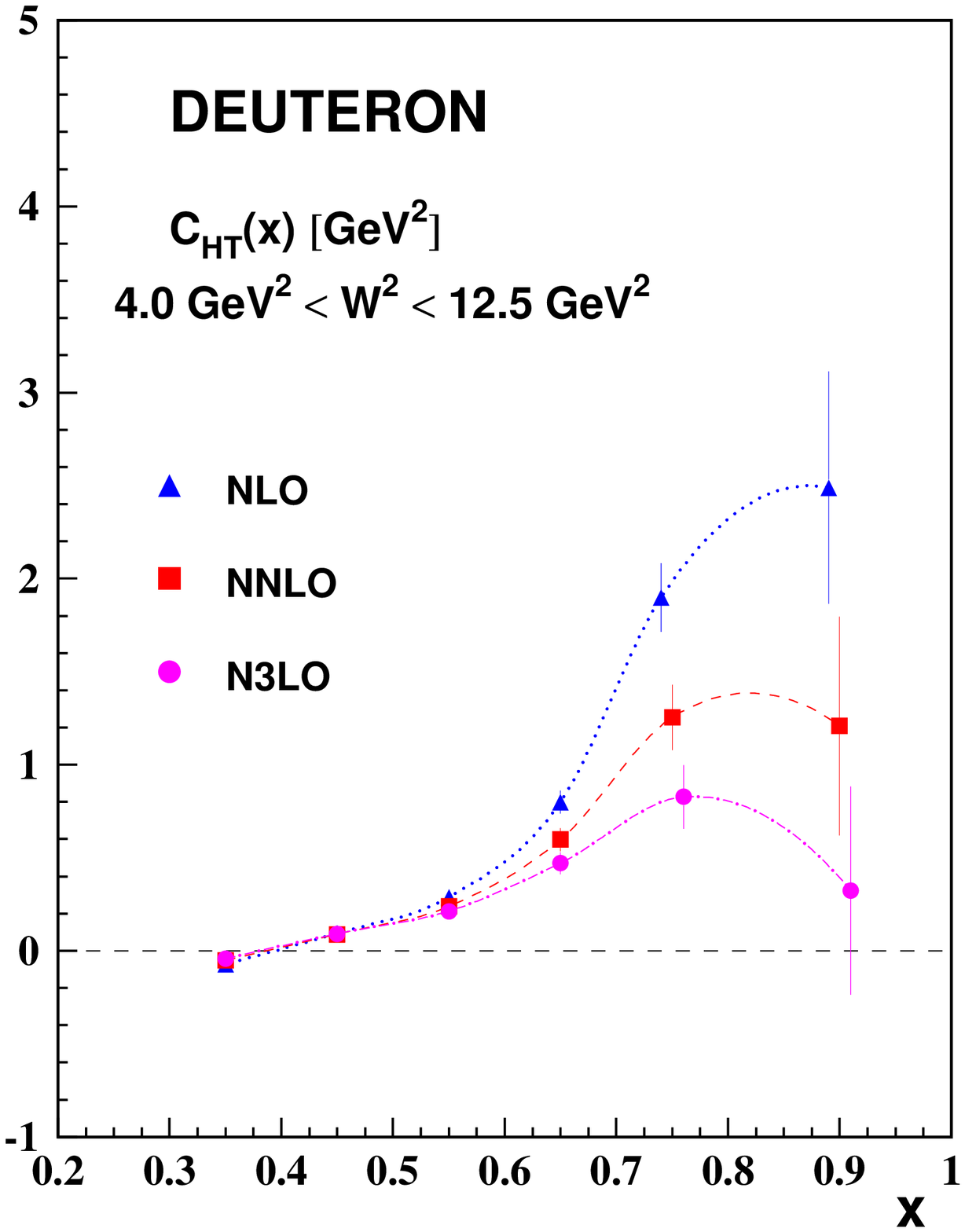,height=7cm,width=7cm}}
\end{center}
\caption[]{
\label{FIG:1}
\small \sf
The empiric higher twist contributions to $F_2^{p,d}(x,Q^2)$ in the valence region, Eq.~(1),~extracted
by calculating the leading twist part at NLO, NNLO, and N$^3$LO$^*$, \cite{BB12a}.}
\end{figure}
%----------------------------------------------------------------------------------------------

%--------------------------------------------------------------------------------------------------
\section{\boldmath $g_2^{\rm tw3}(x,Q^2)$}
%--------------------------------------------------------------------------------------------------

\noindent
Higher twist contributions to the polarized structure function $g_1(x,Q^2)$  have been studied in 
Refs.~\cite{Leader:2009tr,Blumlein:2010rn} in phenomenological approaches aiming on the twist-4 contributions.
However, the structure function $g_2(x,Q^2)$, together with other polarized electro-weak structure functions
\cite{Blumlein:1996tp,Blumlein:1996vs,Blumlein:1998nv}, receives also twist-3 contributions. $g_2(x,Q^2)$ 
obeys the Burkhardt-Cottingham relation \cite{Burkhardt:1970ti}
%--------------------------------------------------------------------------------------------------
\begin{eqnarray}
\int_0^1 dx g_2(x,Q^2) = 0~.
\end{eqnarray}
%--------------------------------------------------------------------------------------------------
Since the Wandzura-Wilczek relation \cite{Wandzura:1977qf} implies, that the first moment of the 
twist-2 part vanishes separately also
%--------------------------------------------------------------------------------------------------
\begin{eqnarray}
\int_0^1 dx g_2^{\rm tw3}(x,Q^2) = 0
\end{eqnarray}
%--------------------------------------------------------------------------------------------------
holds. The errors on the present world data from E143, E155, HERMES and NMC  \cite{DATA} on $g_2(x,Q^2)$ are 
still 
large but yet one may try the fit of a profile in $x$. 
%----------------------------------------------------------------------------------------------
\restylefloat{figure}
\begin{figure}[H]
\begin{center}
\mbox{\epsfig{file=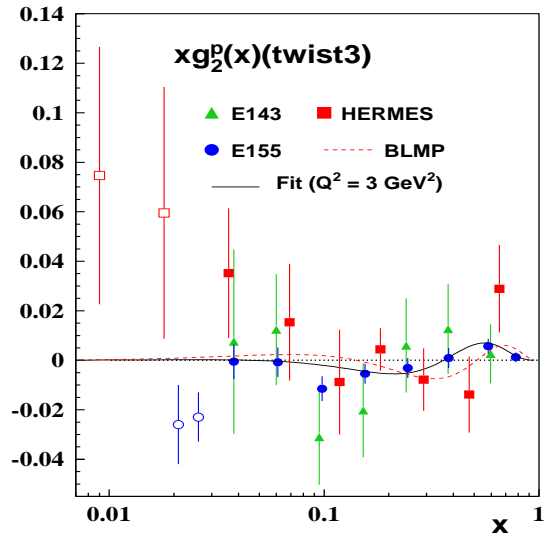,height=7cm,width=7cm}}
\end{center}
\caption[]{
\label{FIG2}
\small \sf
The twist-3 contributions to $g_2(x,Q^2)$ subtracting the twist-2 part according to the Wandzura-Wilczek 
relation \cite{Wandzura:1977qf} using the result of \cite{Blumlein:2010rn} for the twist-2 contribution to 
$g_1(x,Q^2)$ on experimental data from E143, E155, and HERMES \cite{DATA} fitting the shape (\ref{SHA}) (full 
line). Open symbols refer to data in the region $Q^2 < 1~\GeV^2$. The dashed line shows the result of a 
calculation at $Q^2 = 1~\GeV^2$ given in \cite{Braun:2011aw}.}
\end{figure}
%----------------------------------------------------------------------------------------------
In Ref.~\cite{Braun:2011aw} the parameterization
%--------------------------------------------------------------------------------------------------
\begin{eqnarray}
g_2^{\rm tw3}(x) = A \left[ \ln(x) +(1-x) + \frac{1}{2}(1-x)^2\right]
                 +(1-x)^3\left[B - C(1-x) + D(1-x)^2\right]
\label{SHA}
\end{eqnarray}
%--------------------------------------------------------------------------------------------------
has been proposed. Since the data points are measured at different values of $Q^2$ an evolution has
to be performed to a common scale. Furthermore, the target mass corrections \cite{Blumlein:1998nv}
have to be taken into account. In Figure~2 the results of the fit to $g_2^{\rm tw3}(x,Q^2)$ are presented
for $Q^2 = 3~\GeV^2$. We limited the analysis to the region $Q^2 > 1~\GeV^2$. The present errors are still
large and the data of E155 dominate in the fit. We may compare with a theoretical prediction given in 
\cite{Braun:2011aw}. Indeed both results are quite similar.
The twist-3 contribution to the structure function $g_1(x,Q^2)$ can be obtained from that to $g_2(x,Q^2)$
by the integral-relation \cite{Blumlein:1998nv}
%--------------------------------------------------------------------------------------------------
\begin{eqnarray}
g_1^{\rm tw3}(x,Q^2) = \frac{4x^2 M^2}{Q^2} \left[g_2^{\rm tw3}(x,Q^2) - 2 \int_x^1 \frac{dy}{y} g_2^{\rm 
tw3}(y,Q^2) \right]~,
\end{eqnarray}
%--------------------------------------------------------------------------------------------------
cf.~\cite{BB12b}. Due to the large errors of the data the present results are of more qualitative 
character. To study the twist-3 contributions both to the structure functions $g_2(x,Q^2)$ and  
$g_1(x,Q^2)$ in detail, a high luminosity machine, like the planned EIC \cite{Boer:2011fh}, is needed.
%--------------------------------------------------------------------------------------------------
\section{Conclusions}
%--------------------------------------------------------------------------------------------------

\noindent
We performed a re-analysis of the present deep-inelastic world data on proton and deuteron 
targets for the structure function $F_2(x,Q^2)$ in the valence region $x > 0.3$ accounting for remaining 
non-valence tails, which were calculated using the ABKM09 distributions \cite{Alekhin:2009ni}.
We obtain a slightly lower value of $\alpha_s(M_Z^2)$ than in our previous analysis \cite{BBG} 
at N$^3$LO$^*$, however, far within the $1\sigma$ error range. Very stable predictions are obtained going 
from NLO to  N$^3$LO$^*$, both for the valence distribution functions and $\alpha_s(M_Z^2)$. The 
values being obtained for the different sub-sets of experimental data in the present
fit are well in accordance with our global result. We do not confirm the significant differences reported by
MSTW between the SLAC $ep$ and $ed$ data at NNLO \cite{MSTW}. We also disagree with the large value 
of NNPDF \cite{NNPDF} for the BCDMS data at NLO, which also contradicts the corresponding result by MSTW 
\cite{MSTW}. Our results are in agreement with those of the GJR collaboration \cite{GJR} and the singlet 
analyses \cite{Alekhin:2009ni,Alekhin:2012ig}. We obtained an update of the dynamical higher twist contributions 
to $F_2^{p,d}(x,Q^2)$ in the valence region,
which depends on the order to which the leading twist contributions were calculated. The effect stabilizes
including corrections up to N$^3$LO$^*$ in the range $0.3 < x \lsim 0.65$. At larger values of $x$ still higher
order corrections may be needed. A first estimate on the quarkonic twist-3 contributions to the polarized
structure function $g_2(x,Q^2)$ is given in a fit to the available world data on $g_2(x,Q^2)$.
The contributions to $g_1(x,Q^2)$ are obtained by an integral relation, cf. Ref.~\cite{Blumlein:1998nv}.
%--------------------------------------------------------------------------------------------------
\section{Acknowlededgments}

\noindent
For discussions we would like to thank S. Alekhin. This work has been supported in part by DFG 
Sonderforschungsbereich Transregio 9, Computergest\"utzte Theoretische Teilchenphysik, and EU 
Network {\sf LHCPHENOnet} PITN-GA-2010-264564.
% ****************************************************************************
% BIBLIOGRAPHY AREA
% ****************************************************************************

{\raggedright
\begin{footnotesize}
% IF YOU DO NOT USE BIBTEX, USE THE FOLLOWING SAMPLE SCHEME FOR THE REFERENCES
% ----------------------------------------------------------------------------

% ----------------------------------------------------------------------------

% IF YOU USE BIBTEX,
% - DELETE THE TEXT BETWEEN THE TWO ABOVE DASHED LINES
% - UNCOMMENT THE NEXT TWO LINES AND REPLACE 'smith_joe.bib' WITH YOUR
%   FILE(S)

% \bibliographystyle{DISproc}
% \bibliography{smith_joe.bib}
\end{footnotesize}
}

% ****************************************************************************
% END OF BIBLIOGRAPHY AREA
% ****************************************************************************

\end{document}